\documentclass[12pt]{spieman}  
\usepackage{amsmath,amsfonts,amssymb}
\usepackage{graphicx}
\usepackage{setspace}
\usepackage{tocloft}
\usepackage{float}
\usepackage{placeins}
\usepackage{adjustbox}
\usepackage{tablefootnote}
\usepackage{threeparttable}
\usepackage{tikz}
\usepackage{array,multirow,graphicx}
\usepackage{lineno}

\title{Generative Models for Reproducible Coronary Calcium Scoring}

\author[1,2,*]{Sanne G.M. van Velzen}
\author[1]{Bob D. de Vos}
\author[1]{Julia M.H. Noothout}
\author[3]{Helena M. Verkooijen}
\author[2]{Max A. Viergever}
\author[1,4,5,6]{Ivana~I\v{s}gum}
\affil[1]{Department of Biomedical Engineering and Physics, Amsterdam University Medical Centers - location AMC, the Netherlands}
\affil[2]{Image Sciences Institute, Univerisity Medical Center Utrecht and Utrecht University, Utrecht, the Netherlands}
\affil[3]{Imaging Division, Univerisity Medical Center Utrecht, Utrecht, the Netherlands}
\affil[4]{Department of Radiology and Nuclear Medicine, Amsterdam University Medical Centers - location AMC, the Netherlands}
\affil[5]{Amsterdam Cardiovascular Sciences, Amsterdam University Medical Centers, the Netherlands}
\affil[6]{Informatics Institite, Faculty of Science, University of Amsterdam, the Netherlands}

\cftpagenumbersoff{figure}
\cftpagenumbersoff{table} 
\begin{document} 
\maketitle

\begin{abstract}

\noindent \textbf{Purpose:} Coronary artery calcium (CAC) score, i.e. the amount of CAC quantified in CT, is a strong and independent predictor of coronary heart disease (CHD) events. However, CAC scoring suffers from limited interscan reproducibility, which is mainly due to the clinical definition requiring application of a fixed intensity level threshold for segmentation of calcifications. This limitation is especially pronounced in non-ECG-synchronized CT where lesions are more impacted by cardiac motion and partial volume effects. Therefore, we propose a CAC quantification method that does not require a threshold for segmentation of CAC. 
\textbf{Approach:} Our method utilizes a generative adversarial network where a CT with CAC is decomposed into an image without CAC and an image showing only CAC. The method, using a CycleGAN, was trained using 626 low-dose chest CTs and 514 radiotherapy treatment planning CTs. Interscan reproducibility was compared to clinical calcium scoring in radiotherapy treatment planning CTs of 1,662 patients, each having two scans. 
\textbf{Results:} A lower relative interscan difference in CAC mass was achieved by the proposed method: 47\% compared to 89\% manual clinical calcium scoring. The intraclass correlation coefficient of Agatston scores was 0.96 for the proposed method compared to 0.91 for automatic clinical calcium scoring.
\textbf{Conclusions:} The increased interscan reproducibility achieved by our method may lead to increased reliability of CHD risk categorization and improved accuracy of CHD event prediction.

\end{abstract}

{\noindent \footnotesize\textbf{*}Sanne G.M. van Velzen,  \linkable{s.g.m.vanvelzen@amsterdamumc.nl} }

\begin{spacing}{1.2}   

\section{Introduction}
Coronary heart disease (CHD) is among the leading causes of death worldwide \cite{wang2016global}. A clear manifestation of CHD is calcification of the coronary arteries. The amount of coronary artery calcifications (CAC), i.e. coronary calcium score, is a strong and independent predictor of cardiac events, e.g. myocardial infarction or heart failure~\cite{hecht2015coronary, budoff2009coronary}. In CT, CAC is defined as a high density area of $\geq$130 HU~\cite{agatston} in the artery. Using this definition CAC is commonly quantified into CAC volume, mass or Agatston score\cite{agatston}. The Agatston score is used most because it directly indicates a patient's risk of CHD events. 

In a clinical setting the amount of CAC is routinely quantified in non-contrast-enhanced cardiac CT exams with ECG-synchronization. However, other CT scans that show the heart, like chest CT, are increasingly used for quantification of CAC and determination of CVD risk. Such scans are made about ten times more often\cite{de2009projected} than ECG-synchronized cardiac CT and moreover, the CAC scores in these scans correlate well with the CAC scores in cardiac CT\cite{wu2008coronary,budoff2011coronary}. Hence, given its clinical relevance, the guidelines of the Society of Cardiovascular Computed Tomography and the Society of Thoracic Radiology \cite{hecht20172016} recommend to quantify and report CAC on \textit{all} non-contrast CTs showing the heart, including non-ECG-synchronized scans. 
However, the interscan reproducibility of calcium scores is limited. For example, the reported mean interscan variability of Agatston scores in cardiac CT with ECG-synchronization ranges from 15\% to 41\%~\cite{detrano2005coronary, mao2001effect, hoffmann2006evidence, van2003coronary}.
Hence, in non-ECG-synchronized scans the limited interscan reproducibility is further emphasized due to increased influence of partial volume effect and cardiac motion. For instance, Jacobs et al.\cite{jacobs2010coronary} reported an average interscan variability of 71\% for Agatston scores in non-triggered chest CT scans, that led to assignment of different cardiovascular risk category in 24\% of subjects. The compromised insterscan reproducibility limits the applicability calcium quantification in non-ECG-synchronized scans in clinic and for longitudinal studies, in which reproducibility is of key importance to obtain reliable measurements.

\begin{figure}[t]
    \centering
    \includegraphics[height=5cm,trim={1.8cm 3cm 6.5cm 7cm},clip]{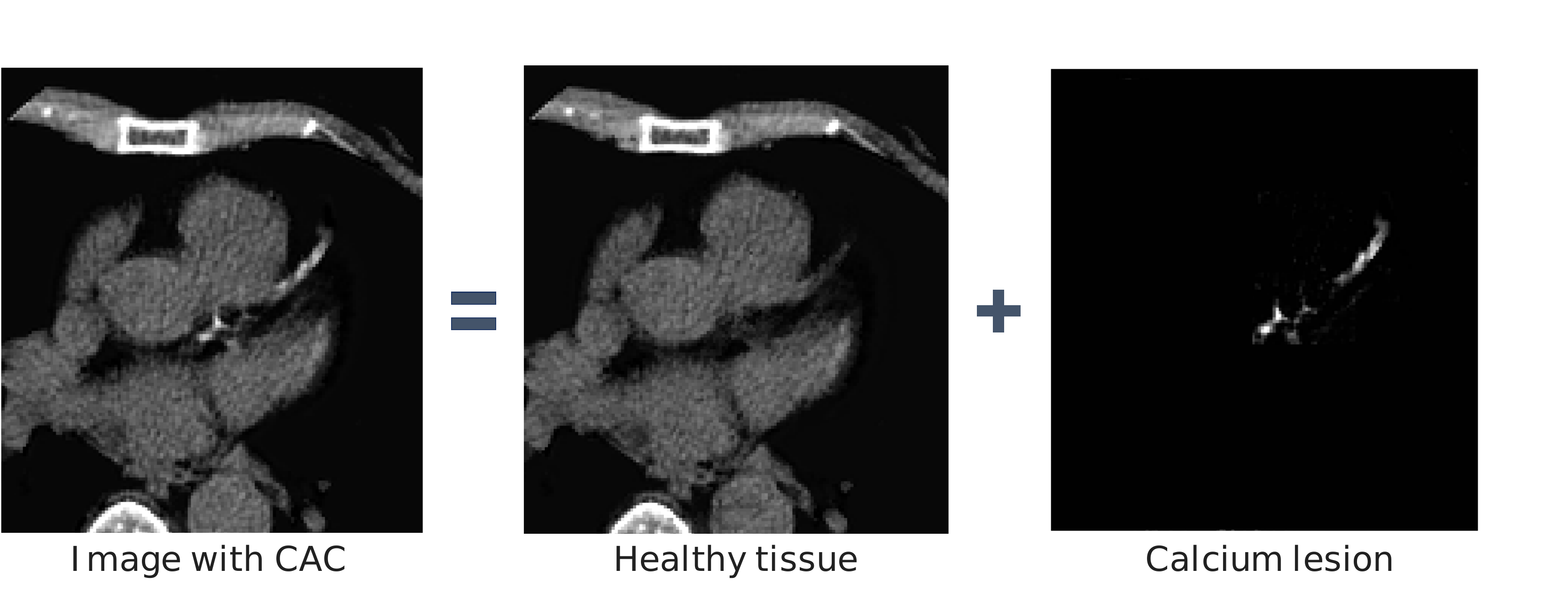}
    \caption{An image containing coronary artery calcium (CAC) is decomposed into an healthy tissue image without CAC and an image containing only CAC.}
    \label{fig:superposition}
\end{figure}

The clinical definition that requires the use of the aforementioned single intensity level threshold at 130 HU has been identified as one of the major causes of the limited reproducibility of CAC quantification\cite{hong2003coronary}, which may lead to under- or overestimation of the amount of CAC. This is especially pronounced in CT exams without ECG-synchronization where visualization of the calcifications is often extremely affected by cardiac motion. As a consequence, the calcifications are blurred or completely remain below the threshold. 
Nevertheless, common to the vast majority of manual and automatic calcium scoring methods is the use of the clinical definition that identifies voxels in the coronary arteries above the 130 HU threshold as CAC in an explicit manner (through segmentations)~\cite{niko,vanvelzen2020deep} or implicit manner (through regression of the CAC score)\cite{devos2019direct, cano2018automated}. 
To address this issue, automatic methods have been proposed that either use automatic adaptive thresholds or omit thresholding for segmentation of calcified lesions. Groen et al.\cite{groen2009threshold} proposed using an adaptive threshold dependent on the maximum intensity value of each CAC lesion. Song et al.~\cite{song2019improved} proposed to adapt the threshold based on the intensity of the background in the vicinity of lesions in combination with deconvolution of the image, using a scanner-specific point spread function to decrease partial volume effect.
Saur et al.~\cite{saur2009accuratum} omitted the intensity threshold for segmentation by employing a mesh-based algorithm that segmented CAC lesions by refining their boundary based on the intensity value profile. 
A different method that circumvents thresholding for segmentation, was proposed by \v{S}prem et al.~\cite{vsprem2018coronary}, who built on work by Dehmeshki et al.~\cite{dehmeshki2007volumetric}. Both used an expectation-maximization algorithm to determine the partial calcium content in each voxel of a CAC lesion and its vicinity.

While the aforementioned methods all show improved interscan reproducibility of CAC quantification, they use the clinically used CAC definition for initial lesion detection, thereby missing lesions below the intensity level threshold. This is particularly a problem with small lesions and scans with motion artefacts, where complete lesions may remain under the detection threshold. Although motion detection \cite{vsprem2018impact} in CAC lesions has been proposed, this does not include correction that would allow quantification of such CAC lesions. 

To address the limited interscan reproducibility of CAC detection and quantification, we propose a method that does not depend on the current clinically used definition of CAC that requires intensity-level thresholding. Instead, we define CAC as the difference between an image visibly containing CAC and a healthy tissue image. Our method takes an image with CAC and decomposes it into an image without CAC and a CAC-map, i.e. an image showing only CAC (Figure \ref{fig:superposition}). The CAC-map comprises the location of lesions and the grey values that indicate the attenuation in the CT scan due to CAC lesions. Due to this new definition of CAC lesions, our method can detect lesions that are completely - and partly - below the clinically used threshold.

Several earlier studies proposed to use an autoencoder for generating healthy counterparts of images with pathology in an unsupervised manner\cite{chen2018unsupervised,sato2018primitive,pawlowski2018unsupervised}. Subsequently, the pixel-wise intensity difference between the original image and the reconstruction was used to e.g. detect brain lesions in MRI scans~\cite{chen2018unsupervised}, brain CT~\cite{pawlowski2018unsupervised}, or detected anomalies in head CT~\cite{sato2018primitive}.
Alternative approaches used generative adversarial networks (GAN) for translating an image containing pathologies to a healthy image. Baur et al. \cite{baur2018deep} and Seah et al. ~\cite{seah2019chest} used conditional GANs to generate healthy images and used the difference with the input image for multiple sclerosis lesion segmentation, and visualizing features of congenital heart failure in chest X-rays, respectively. Furthermore, Baumgartner et al. \cite{baumgartner2018visual} proposed to use a Wasserstein GAN to directly generate a difference map that indicates changes in the brain related to Alzheimer's disease. The map is added to the input image to create a healthy brain image from an image containing Alzheimer's disease. Tang et al. \cite{tang2021disentangled} proposed to predict both an healthy image and a difference map that indicates disease using a disentangled conditional GAN.
To tackle the instability and mode collapse problems that often arise during training of GANs, Zhu et al.~\cite{zhu2017unpaired} proposed to use a cycle-consistency loss in their CycleGAN for image-to-image translation using natural images. Sun et al.~\cite{sun2020adversarial} proposed to use a CycleGAN implementation for translating images containing lesions to healthy images. Subsequently, the pixel-wise intensity difference was used to detect brain lesions in MRI and liver tumors in CT scans.

To decompose an image with CAC into an image without CAC and CAC-map, we employ a CycleGAN to translate images between the \textit{containing CAC} domain and \textit{not containing CAC} domain. To our knowledge, our method is the first to use a generative model for accurate segmentation of CAC lesions. The method builds on our preliminary work ~\cite{vanvelzen2020coronary}, in which we presented a semi-automatic method for segmentation of CAC, that analyzed the vicinity of manually identified CAC lesions. The method was trained using clinical calcium scores as reference and hence, was not able to detect lesions below the threshold. In the current work, we extend the method to perform fully automatic CAC detection and quantification, including CAC lesions below the standard detection threshold. The method is aimed at scans where the low interscan reproducibility is most dire and therefore, experiments were performed in non-ECG-synchronized scans, with and without breathing motion, that often contain excessive cardiac motion artefacts and have a low spatial resolution. To simplify the challenging task and allow the method to focus on the region of interest only, CAC quantification was performed in the heart slices containing CAC. The method is developed and evaluated using 4,038 CT scans of 2,276 breast cancer patients and 626 CT scans of lung screening participants.
Moreover, we show that the here proposed method outperforms the gold standard for calcium scoring in the clinic and a state-of-the-art automatic calcium scoring method in terms of interscan reproducibility of CAC quantification and detection of visible lesions.



\section{Data}
\label{section:data}
\begin{figure}[t]
    
    \centering
    \includegraphics[height=6cm,trim={2.5cm 7.5cm 8cm 6.5cm},clip]{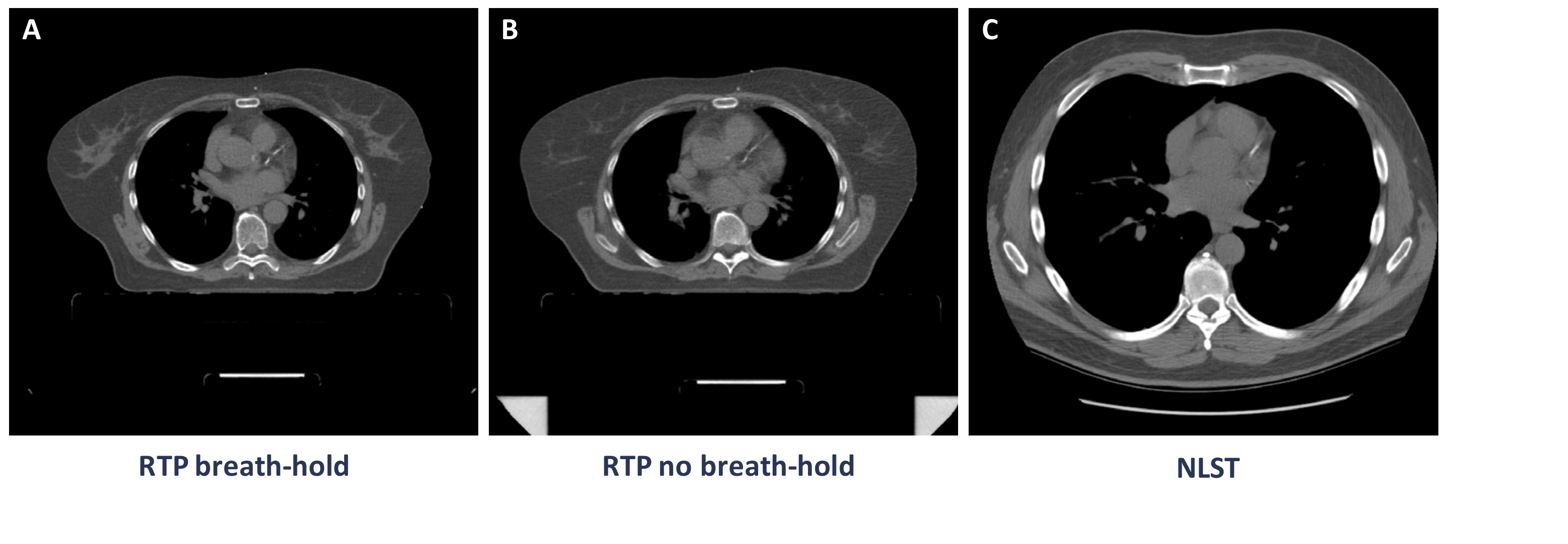}
    \caption{Examples of two RTP scans of one breast cancer patient, one with breath-hold (A) and one without breath-hold (B), and an NLST scan of one participant (C). }
    \label{fig:ex_scans}
\end{figure}

We confirm that all methods were carried out in accordance with relevant guidelines and regulations. We included 4,038 radiotherapy treatment planning (RTP) CTs of 2,276 breast cancer patients\cite{emaus2019bragatston}. In this set 1,762 patients had two scans made on the same day: one with and one without breath-hold (Figure \ref{fig:ex_scans}, A and B). Therefore, this set allows evaluation of CAC scoring interscan reproducibility. The remaining 514 patients had one CT scan made without breath-hold. Scans were acquired in the University Medical Center Utrecht with a Philips Brilliance Big Bore scanner. Intravenous (IV) contrast was not induced and the acquisition was not ECG-synchronized. The CTs were acquired with 120 kVp and reconstructed to 0.92-1.37 mm in-plane resolution and 3.0 mm slice thickness and increment. This study was approved and the need for informed consent was waived by the institutional review board of the University Medical Center Utrecht, the Netherlands.

Additionally, we included 626 low-dose chest CT images of 626 subjects included in the National Lung Screening Trail (NLST, Figure \ref{fig:ex_scans}C). Scans were acquired in 29 different hospitals, on scanners from all major vendors \cite{niko ,national2011reduced}. Tube voltage was either set to 120 or 140 kVp, depending on the subjects weight. IV contrast was not induced and acquisition was not ECG-synchronized. Images were reconstructed to 0.49-0.98 mm in-plane resolution and 1.0-2.5 mm slice thickness and 0.6-2.5 mm increment.

Table \ref{tab:data} lists the used data sets, availability of reference annotations and describes the purpose of each data set. For development and training of the method, RTP CT scans of 514 patients with a single CT ($RTP_{train}$ and $RTP_{train HS}$) and all 626 NLST scans ($NLST_{train}$) were used. The latter were used to enhance training with a diverse set of scans of heavy smokers with high CAC burden. In method development, a validation set of 100 patients with two available RTP scans (200 scans) was used ($RTP_{val}$). We divided the test scans into two sets with patient scan pairs: $RTP_{test1}$, a set consisting of scans of 119 patients totalling 238 RTP CTs with manually segmented CAC allowing detailed evaluation of calcium scoring; and $RTP_{test2}$, a set consisting of scans of 1,543 patients of in total 3,086 RTP CTs without reference CAC scores, used for assessment of interscan reproducibility.


\begin{figure}[t]
    \centering
    \includegraphics[height=8cm,trim={1.8cm 3cm 8cm 2cm},clip]{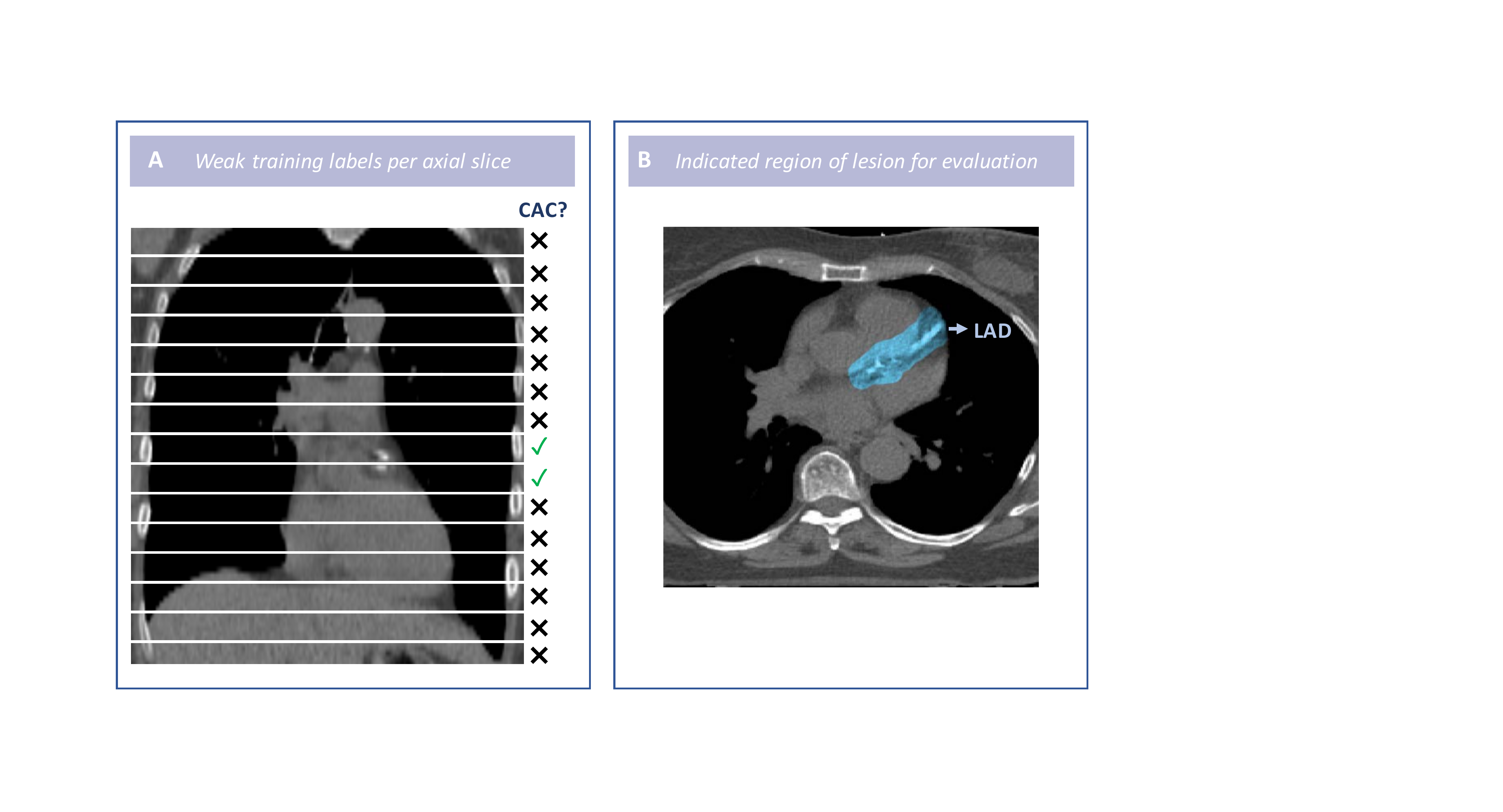}
    \caption{Schematic example of the weak training labels per axial slice indicating whether CAC is present in the slice (A). Indicated region of the lesion with assigned coronary artery label (B). }
    \label{fig:labels}
\end{figure}

\begin{table*}[h!]

\footnotesize
\caption{A. Data sets, consisting of low-dose CT scans from the National Lung Screening Trial (NLST) or radiotherapy treatment planning (RTP) CT scans, used for development ($NLST_{train}, RTP_{train}, RTP_{train HS}, RTP_{val}$) and evaluation ($RTP_{test HS}, RTP_{test1}, RTP_{test lesions}, RTP_{test2}$) of the presented method. Scans are made with (BH) or without deep inspiration breath-hold (nonBH). B. Availability of manual reference annotations are indicated per set for heart segmentations (Heart Segm.), labels per axial slice indicating presence of CAC as a binary decision assessmed visually (Figure \ref{fig:labels}A) and availability of coarse segmentations of lesion regions (Lesion Regions, Figure \ref{fig:labels}B). C. The way the data is used in the experiments is listed as training (Tr), validation (Val) and testing (Test) for the different networks (Heart Segmentation, CAC Classification, Calcium CycleGAN). }
\label{tab:data}

\centering
\setlength{\tabcolsep}{5pt}
\begin{threeparttable}[b]

\begin{tabular}{llccccccccc}
\hline
\textbf{A} & \multicolumn{10}{c}{\textbf{Scan information}}                                                                                                                                                                                                                                  \\ \hline
           & \multicolumn{1}{l}{Scan Type} & \multicolumn{3}{c}{Patients}                                                      & \multicolumn{3}{c}{BH}       & \multicolumn{3}{c}{nonBH}     \\ \hline
\multicolumn{2}{l}{\textbf{Development}}  & \multicolumn{3}{l}{}                                                              & \multicolumn{3}{l}{}                                                         & \multicolumn{3}{c}{}                                                          \\
           & $NLST_{train}$                         & \multicolumn{3}{c}{626}                                                           & \multicolumn{3}{c}{626}                                                      & \multicolumn{3}{c}{}                                                          \\
           & $RTP_{train}$                          & \multicolumn{3}{c}{514}                                                           & \multicolumn{3}{c}{}                                                         & \multicolumn{3}{c}{514}                                                       \\
           & $RTP_{train HS}$                 & \multicolumn{3}{c}{120}                                                           & \multicolumn{3}{c}{}                                                         & \multicolumn{3}{c}{120}                                                       \\
           & $RTP_{val}$                          & \multicolumn{3}{c}{100}                                                           & \multicolumn{3}{c}{100}                                                      & \multicolumn{3}{c}{100}                                                       \\
\multicolumn{2}{l}{\textbf{Evaluation}}   & \multicolumn{3}{c}{}                                                              & \multicolumn{3}{c}{}                                                         & \multicolumn{3}{c}{}                                                          \\
           & $RTP_{test HS}$                          & \multicolumn{3}{c}{79}                                                            & \multicolumn{3}{c}{}                                                         & \multicolumn{3}{c}{79}                                                        \\
           & $RTP_{test1}$                          & \multicolumn{3}{c}{119}                                                           & \multicolumn{3}{c}{119}                                                      & \multicolumn{3}{c}{119}                                                       \\
           & $RTP_{test lesions}$                 & \multicolumn{3}{c}{34}                                                            & \multicolumn{3}{c}{34}                                                       & \multicolumn{3}{c}{34}                                                        \\
           & $RTP_{test2}$                          & \multicolumn{3}{c}{1,543}                                                         & \multicolumn{3}{c}{1,543}                                                    & \multicolumn{3}{c}{1,543}                                                     \\ \hline
\textbf{B} & \multicolumn{10}{c}{\textbf{Reference Annotations}}                                                                                                                                                                                                                             \\ \hline
           &                              & \multicolumn{3}{c}{\begin{tabular}[c]{@{}c@{}}Heart \\ Segmentation\end{tabular}} & \multicolumn{3}{c}{\begin{tabular}[c]{@{}c@{}}Per Slice \\ CAC\end{tabular}} & \multicolumn{3}{c}{\begin{tabular}[c]{@{}c@{}}Lesion \\ Regions\end{tabular}} \\ \hline
\multicolumn{2}{l}{\textbf{Development}}  & \multicolumn{3}{c}{}                                                              & \multicolumn{3}{c}{}                                                         & \multicolumn{3}{c}{}                                                          \\
           & $NLST_{train}$                         & \multicolumn{3}{c}{}                                                              & \multicolumn{3}{c}{\checkmark}                                                    & \multicolumn{3}{c}{}                                                          \\
           & $RTP_{train}$                          & \multicolumn{3}{c}{}                                                              & \multicolumn{3}{c}{\checkmark}                                                    & \multicolumn{3}{c}{}                                                          \\
           & $RTP_{train HS}$                 & \multicolumn{3}{c}{\checkmark}                                                         & \multicolumn{3}{c}{}                                                         & \multicolumn{3}{c}{}                                                          \\
           & $RTP_{val}$                          & \multicolumn{3}{c}{}                                                              & \multicolumn{3}{c}{\checkmark}                                                    & \multicolumn{3}{c}{}                                                          \\
\multicolumn{2}{l}{\textbf{Evaluation}}   & \multicolumn{3}{c}{}                                                              & \multicolumn{3}{c}{}                                                         & \multicolumn{3}{c}{}                                                          \\
           & $RTP_{test HS}$                          & \multicolumn{3}{c}{\checkmark}                                                         & \multicolumn{3}{c}{}                                                         & \multicolumn{3}{c}{}                                                          \\
           & $RTP_{test1}$                          & \multicolumn{3}{c}{}                                                              & \multicolumn{3}{c}{\checkmark}                                                    & \multicolumn{3}{c}{}                                                          \\
           & $RTP_{test lesions}$                 & \multicolumn{3}{c}{}                                                              & \multicolumn{3}{c}{\checkmark}                                                    & \multicolumn{3}{c}{\checkmark}                                                     \\
           & $RTP_{test2}$                          & \multicolumn{3}{c}{}                                                              & \multicolumn{3}{c}{}                                                         & \multicolumn{3}{c}{}                                                          \\ \hline
\textbf{C} & \multicolumn{10}{c}{\textbf{Purpose}}                                                                                                                                                                                                                                           \\ \hline
           &                              & \multicolumn{3}{c}{\begin{tabular}[c]{@{}c@{}}Heart\\ Segmentation\end{tabular}}  & \multicolumn{3}{c}{\begin{tabular}[c]{@{}c@{}}CAC\\ Classification\end{tabular}}                                       & \multicolumn{3}{c}{\begin{tabular}[c]{@{}c@{}}Calcium\\ CycleGAN\end{tabular}}                                          \\
           &                              & Tr                        & Val                      & Test                    & Tr                    & Val                     & Test                    & Tr                     & Val                     & Test                    \\ \hline
\multicolumn{2}{l}{\textbf{Development}}  & \textbf{}                    &                          &                         &                          &                         &                         &                           &                         &                         \\
           & $NLST_{train}$                         &                              &                          &                         & \checkmark                    &                         &                         & \checkmark                     &                         &                         \\
           & $RTP_{train}$                          &                              &                          &                         & \checkmark                    &                         &                         & \checkmark                     &                         &                         \\
           & $RTP_{train HS}$                 & \checkmark                        & \checkmark                    &                         &                          &                         &                         &                           &                         &                         \\
           & $RTP_{val}$                          &                              &                          &                         &                          & \checkmark                   &                         &                           & \checkmark                   &                         \\
\multicolumn{2}{l}{\textbf{Evaluation}}   & \textbf{}                    &                          &                         &                          &                         &                         &                           &                         &                         \\
           & $RTP_{test HS}$                          &                              &                          & \checkmark                   &                          &                         &                         &                           &                         &                         \\
           & $RTP_{test1}$                          &                              &                          &                         &                          &                         & \checkmark                   &                           &                         & \checkmark                   \\
           & $RTP_{test lesions}$                 &                              &                          &                         &                          &                         & \checkmark                   &                           &                         & \checkmark                   \\
           & $RTP_{test2}$                          &                              &                          &                         &                          &                         & \checkmark                   &                           &                         & \checkmark                   \\ \hline
\end{tabular}

\end{threeparttable}

\end{table*}

To allow training, all training CTs, validation CTs and CTs from $RTP_{test1}$ were manually labelled without using a threshold by an observer with $>$3 years experience in calcium scoring. Because voxelswise annotation without a threshold is practically infeasible, as the exact outline of the CAC is not known, weak labels were used to train the methods. Therefore, the observer labelled axial slices of the scans as \textit{containing CAC} or \textit{not containing CAC}, based on visual assessment (Figure \ref{fig:labels}A). Moreover, for a subset of 199 RTP scans of the training set ($RTP_{train HS}$ and $RTP_{test HS}$), manual segmentations of the heart were available from clinical practice.

To define reference standard for CAC, lesions were annotated and assigned coronary artery label in $RTP_{test1}$. Because accurate voxel-level manual annotation without threshold is hardly feasible, regions containing lesions were indicated in the scans, by rough manual voxel-painting without a threshold, with a brush size of 1 cm (Figure \ref{fig:labels}B). This was only done in the 34 scan pairs in which both scans had CAC, because pairs without CAC and visibly discordant pairs are not informative for measuring interscan reproducibility. Visibly discordant pairs are pairs in which in one scan CAC is visible and in the second scan it is not. Next, to allow comparison with calcium scoring used in clinic, manual annotations following the clinical definition from a previous study for $RTP_{test1}$ were available~\cite{vanvelzen2020deep}.

\begin{figure}[ht!]
    \centering
    \includegraphics[height=9cm,trim={1cm 8cm 15cm 1.5cm},clip]{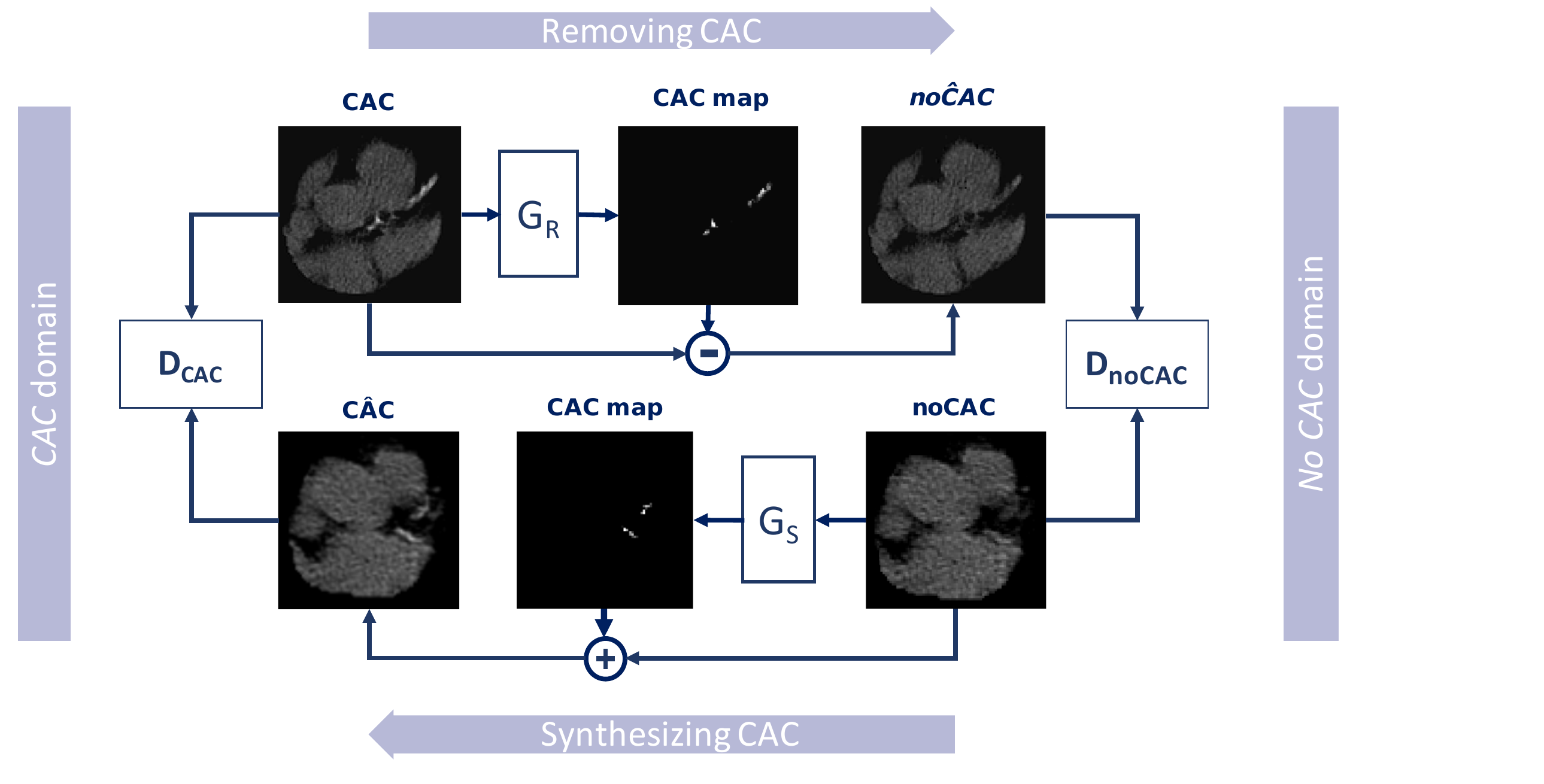}
    \caption{Schematic overview of the proposed CycleGAN that translates the images from the domain of images with CAC (CAC domain) to the domain of images without CAC (No CAC domain) and back. In the pathway that removes CAC, a CAC map is predicted by the generator (G\textsubscript{R}) and subtracted from the image to obtain a synthetic image without CAC. In the pathway that synthesizes CAC the CAC map predicted by the generator (G\textsubscript{S}) is added to the image. Synthetic images are compared to real examples by discriminators (D).}
    \label{fig:cyclegan}
\end{figure}

\section{Method}
\label{section:method}
We propose an automatic method that detects and segments CAC by decomposing a CT slice into an image without CAC and an image showing a CAC map, i.e. only CAC. By defining CAC as the difference between an image visibly containing CAC and a healthy tissue image, the method allows segmentation of whole lesions, including parts below the current clinical definition applying 130 HU threshold, leading to increased interscan reproducibility of CAC quantification. 

The need for improved interscan reproducibility of CAC quantification is highest in non-ECG-synchronized scans. However, non-ECG-synchronized scans are typically made for other purposes than calcium scoring, thus, the field of view in these scans is highly variable. Therefore, before generating the CAC-map, we define the region of interest to simplify the CAC detection task by initial heart segmentation, and subsequent identification of CT slices with CAC.

\begin{table*}[]
\footnotesize
\caption{Architecture of the Heart Segmentation CNN, CAC Classification CNN and the Generators in the proposed CycleGAN}
\label{tab:resnets}

\centering
\setlength{\tabcolsep}{5pt}
\begin{threeparttable}[b]
\begin{tabular}{lllll}
      &                              & \textbf{Heart Segmentation CNN} & \textbf{CAC Classification CNN} & \textbf{CAC-map generators} \\ 
      & & & & \textbf{CycleGAN}\\\hline
\multicolumn{2}{l}{\textbf{General}} &                             &                         &                                  \\
      & Dimensions                   & 3D                          & 2D                      & 2D                               \\
      & Input size                   & 64x64x64 voxels             & 224x224 voxels          & 224x224 voxels                   \\
      & Type of padding (layers)     & Zero (All)                  & Reflection (All)        & Reflection (All)                 \\ \hline
\multicolumn{5}{l}{\textbf{Architecture (kernel, output channels, stride)} }                                                    \\
      & 1st Layer                    & Conv (7x7x7, 16)            & Conv (7x7, 64)          & Conv (7x7, 64)                   \\
      & 2nd Layer                    & Conv (3x3x3, 32, 2)         & Conv (3x3, 128, 2)      & Conv (3x3, 128, 2)               \\
      & 3rd Layer                    & Conv (3x3x3, 64, 2)         & Conv (3x3, 256, 2)      & Conv (3x3, 256, 2)               \\
      & ResNet Blocks                & 9 Blocks (3x3x3, 64, 1)          & 6 Blocks (3x3, 256, 1)     & 6 Blocks (3x3, 256, 1)              \\
      & 3rd last Layer               & TConv (3x3x3, 32, 2)       & -    & TConv (3x3, 128, 2)                                \\
      & 2nd last Layer               & TConv (3x3x3, 16, 2)       & Global Average Pool     &  TConv (3x3, 64, 2)             \\
      & Last Layer                   & Conv (7x7x7, 2)             &  Dense Layer (256x2 channels)        & Conv (7x7x7, 1)          \\
      & Activation last Layer        & Sigmoid                     &  Softmax                &  Sigmoid                         \\ \hline
\end{tabular}
\begin{tablenotes}
\item Conv = Convolution layer; TConv = Transposed convolution layer; All convolution layers, except the last layer, are followed by a batch normalization layer and rectifier linear unit activation. Resnet Blocks consist of 2 convolution layer pairs with a residual connection.
\end{tablenotes}
\end{threeparttable}

\end{table*}

Heart segmentation is performed with a 3D convolutional neural network (CNN), which, for efficiency, analyses the full image in a patch based manner. The network architecture is based on ResNet~\cite{he2016deep} and is detailed in Table \ref{tab:resnets}, left. The Dice loss is used during training and Adam optimization algorithm ($lr=0.001$) is used for optimization. During inference, overlapping patches were analysed and the results were averaged for final prediction.

We classify each slice in the heart mask according to the presence of visible CAC with a 2D CNN with ResNet~\cite{he2016deep} based architecture (Table \ref{tab:resnets}, middle). The CNN analyses axial image slices centered and cropped around the heart. For cropping, the center of mass of the heart segmentation is used. The network is trained with per slice labels indicating the presence of CAC, using the cross-entropy loss and Adam optimization algorithm ($lr=0.001$).

For the identified slices, we generate a CAC-map. Manually identifying voxels with visible CAC to directly predicts a CAC-map with a CNN, would be hardly feasible. Hence, we propose to obtain a CAC-map using weakly supervised labels, i.e. labels indicating whether CAC is present in the image. In our approach a CAC-map is obtained by subtracting a generated healthy tissue image from an image containing CAC. Because it is not possible to obtain paired CT scans of the same patient, where one would contain CAC and the other not, a paired approach for image generation is not feasible. Therefore, an unpaired approach is used to generate synthetic $\hat{noCAC}$ images from images containing CAC. For this, images are assigned to two domains based on weak labels: one domain consisting of images containing visible CAC and the other domain containing images without visible CAC. Hence, to generate a healthy tissue image for subtraction, a 2D CycleGAN\cite{zhu2017unpaired} for unpaired image-to-image translation is used to translate images from one domain to another. The CycleGAN is a type of GAN~\cite{goodfellow2014generative} that improves in training stability and performance over the traditional and conditional GAN, by adding cycle-consistency~\cite{zhu2017unpaired}.

The CycleGAN consists of two generators $G$, and two discriminators $D$ that compete with each other. The generators are CNNs that convert an input image from either domain to a realistic image from the other domain. In our task the images are divided into the \textit{CAC} domain that contains images with CAC and the \textit{noCAC} domain that contains images without CAC. The discriminator CNNs compare the synthesized images against \textit{real} examples from the target domain. In contrast to a conventional CycleGAN, that directly translates an input image to an image of the target domain, we use an adjusted CycleGAN (Figure \ref{fig:cyclegan}) that adds an extra step to minimize data hallucination: in the removing pathway, images with CAC are translated to the \textit{noCAC} domain by predicting a CAC map and subtracting it from the input image: $\hat{noCAC} = CAC - G_R(CAC)$. Hence, the CAC map $G_R(CAC)$ comprises both the location of CAC lesions and grey values that indicate the attenuation in the CT scan due to CAC lesions. In the synthesizing pathway the CAC map $G_S(noCAC)$ is added to the input image to generate a synthetic image containing CAC from an image without CAC:  $\hat{CAC} = noCAC + G_S(noCAC)$. In this pathway the generator learns to synthesize realistic lesions based on probable lesion locations and severity inferred from the training data. In this work the synthesizing pathway is solely utilized for training stability. The generator in the pathway that removes CAC, i.e. $G_R$, is used to obtain the CAC map and, subsequently, the CAC map can be used to quantify the amount of CAC in the image.

The generators used in the CycleGAN for generating the CAC map have a ResNet-based architecture~\cite{he2016deep}, with 6 ResNet blocks using ReLU activation and no downsampling in between blocks (Table \ref{tab:resnets}, right). To ensure the CAC maps only contain positive values, the generators have a sigmoid activation in the last layer. For the discriminator networks we use PatchGANs, which classify 70x70 overlapping image patches as real or synthetic, described in detail by \cite{isola2017image}. The PatchGANs are able to capture high-frequency structure and have fewer parameters than a full-image discriminator and are therefore, easier to train~\cite{isola2017image}. The patch-level discriminators are applied to the image in a fully convolutional manner, averaging all responses to provide the output of the discriminator. 

The system is trained with the loss terms on the input and synthetic images that are traditionally used for training a CycleGAN\cite{zhu2017unpaired}:  Adversarial loss (Equation \ref{eq:adv}), Cycle-consistency loss (Equation \ref{eq:cyc}) and Identity loss (Equation \ref{eq:id}). The Identity loss trains \( G_{R} \) to predict an empty CAC-map when it is presented with an image without CAC and vice versa for \( G_{S} \). Additionally, a Sparsity (L1) loss (Equation \ref{eq:sp}) on the CAC map is used to encourage the generator to predict sparse CAC maps, giving the full objective:

\setlength{\abovedisplayskip}{-15pt}
\setlength{\belowdisplayskip}{0pt}
\setlength{\abovedisplayshortskip}{0pt}
\setlength{\belowdisplayshortskip}{0pt}

\begin{equation}
\label{eq:full}
\textbf{L}  =  \textbf{L}_{Adversarial} + \lambda\textbf{L}_{Cycle} + \alpha\textbf{L}_{Identity} + \beta\textbf{L}_{Sparsity}
\end{equation}

Where

\begin{multline}
\label{eq:adv}
\textbf{L}_{Adversarial}  =  \mathbb{E}_{CAC \sim P_{data(CAC)}}[\log \textbf{D}_{CAC}(CAC)] + \mathbb{E}_{noCAC \sim P_{data(noCAC)}}[\log (1-\textbf{D}_{CAC}(\hat{CAC}))] \\ 
+ \mathbb{E}_{noCAC \sim P_{data(noCAC)}}[\log \textbf{D}_{noCAC}(noCAC)] \\
+ \mathbb{E}_{CAC \sim P_{data(CAC)}}[\log (1-\textbf{D}_{noCAC}(\hat{noCAC}))]
\end{multline}

\begin{multline}
\label{eq:cyc}
\textbf{L}_{Cycle}  =  \mathbb{E}_{CAC \sim P_{data(CAC)}}||\textbf{G\textsubscript{S}}(\textbf{G\textsubscript{R}}(CAC)) - CAC||_1 \\
+ \mathbb{E}_{noCAC \sim P_{data(noCAC)}}||\textbf{G\textsubscript{R}}(\textbf{G\textsubscript{S}}(noCAC)) - noCAC||_1 
\end{multline}

\begin{multline}
\label{eq:id}
\textbf{L}_{Identity}  =  \mathbb{E}_{noCAC \sim P_{data(noCAC)}}||\textbf{G\textsubscript{R}}(noCAC)||_1 + \mathbb{E}_{CAC \sim P_{data(CAC)}}||\textbf{G\textsubscript{S}}(CAC)||_1 
\end{multline}

\begin{multline}
\label{eq:sp}
\textbf{L}_{Sparsity}  =  \mathbb{E}_{CAC \sim P_{data(CAC)}}||\textbf{G\textsubscript{R}}(CAC)||_1 + \mathbb{E}_{noCAC \sim P_{data(noCAC)}}||\textbf{G\textsubscript{S}}(noCAC)||_1 
\end{multline}

To make the method more robust to noise, during training images are augmented by reducing or amplifying noise. To reduce the noise, images are smoothed with a Gaussian filter with $\sigma = 0.5$ pixels. To amplify noise, a noise image is obtained by subtracting the smoothed image from the original input image that is, subsequently, added to the input image. Moreover, during training random cropping and rotation are used for augmentation.

Despite the sparsity loss, the generated CAC-map may contain low levels of noise, because of excessive image noise. Noise in the CAC map may lead to unrealistically low values for soft tissue in the synthetic noCAC image. Therefore, we mask voxels in the CAC-map that lead to a value of $<$ -10 HU in the synthetic noCAC image, as those are likely noise voxels in the CAC-map.

Finally, CAC was quantified by CAC pseudo-mass and adjusted Agatston score. The pseudo-mass was calculated by multiplying the sum of HU values of the voxels in the lesions by the voxel volume.
The adjusted Agatston score was calculated by multiplying the area of a lesion with a density score (1: $<$199 HU, 2: 200–299 HU, 3: 300–399 HU, 4: $>$399 HU) determined by the maximum density of the area, and summing over axial slices. Please note that the adjusted Agatston score includes segmented (parts of) lesions below the clinically used threshold. For comparison, the conventional Agatston score was calculated based on segmentations obtained using the clinical definition, which uses the following density scores: 0: $<$130 HU, 1: 130-199 HU, 2: 200–299 HU, 3: 300–399 HU, 4: $>$399 HU. Please note that the adjusted Agatston score is equal to the conventional Agatston score when CAC lesions are segmented using the 130 HU threshold according to the clinical definition of CAC.

\section{Evaluation}
\label{section:evaluation}
Detection of CAC lesions was evaluated using accuracy, sensitivity, rate of false positive scans and F1-score, between the automatic and reference labels that indicated visible CAC. 

The reproducibility of the CAC score was evaluated in scan pairs, using absolute relative differences of quantified CAC pseudo masses. To measure the interscan agreement of adjusted Agatston scores, the two-way random intraclass correlation coefficient (ICC) for absolute agreement was computed and Bland Altman plots with 95\% limits of agreement were examined\cite{sevrukov2005serial}. Moreover, patients were assigned one of four CHD risk categories (I: 0–10, II: 11–100, III: 101–400, IV: $<$400) based on their adjusted Agatston score. Agreement between risk categories of each patient determined from the patient's two scans was evaluated using Cohen's linearly weighted kappa ($\kappa$).



\section{Experiments and Results}
\label{sec:exp}
We evaluate the detection performance of visible CAC and the interscan reproducibility of the proposed method. Moreover, we compare the performance of our method with two clinical calcium scoring methods that use the standard clinically used CAC definition for segmentation: manual calcium scoring and automatic calcium scoring with a validated, state-of-the-art deep learning approach\cite{vanvelzen2020deep}. Finally, we perform an ablation study to evaluate the different components of the method.

Prior to heart segmentation all scans were resampled to 1.5 mm in-plane resolution for standardization across the data set and 3.0 mm slice spacing, the slice spacing of manual reference annotations. The heart segmentation CNN was trained using $RTP_{train HS}$: a subset of 120 radiotherapy treatment planning CT scans of $RTP_{train}$, for which reference heart segmentations were available. Of this set, 100 scans were used for training and 20 scans for validation. $RTP_{test HS}$ was used for evaluation (see Table \ref{tab:data}). The CNN was trained for 250,000 iterations with batches of 10 patches of 64x64x64 voxels. Batches were balanced for presence of the heart during training. After training, the heart segmentation CNN achieved a median Dice score of 0.95 (interquartile range 0.94-0.96), a median Haussdorf distance of 15 mm (interquartile range 11 - 21 mm) and a median asymmetric surface distance of 1.6 mm (interquartile range 1.2 - 2.2 mm) over all test images in $RTP_{test HS}$.

After the heart segmentation, images were resampled to 1.0 mm in-plane resolution, which was the average resolution of the training set, and 1.5 mm slice spacing, for comparison with clinical calcium scoring\cite{vanvelzen2020deep}. Moreover, slices were clipped between \mbox{-50 HU} and 950 HU, for optimal contrast between soft tissue and CAC, and scaled between 0 and 1. The training set of 626 low dose chest CTs ($NLST_{train}$) and 315 radiotherapy treatment planning CTs ($RTP_{train}$) was used, with in total 65,675 2D slices containing the heart. Given that the method is aimed for analysis of RTP CT, the validation set of 200 such scans of 100 breast cancer patients (100 pairs) was used ($RTP_{val}$, Table \ref{tab:data}). During training manually defined reference labels indicating the presence of CAC in axial CT slices were used. 
The slice classification CNN was trained for 1,500,000 iterations using batches of 20 axial CT slices of 224x224 voxels. Batches were randomly sampled and balanced for presence of visible CAC. After training, the slice classification CNN achieved an per slice accuracy of 0.93, with a sensitivity of 0.78 for visible CAC and a false positive rate of 0.06, on all 17,450 CT slices from $RTP_{test1}$ containing the heart. On average 9.9\% of the slices containing the heart was analysed by the CycleGAN.

\begin{figure}
    \centering
    \includegraphics[height=6.5 cm,trim={2.9cm 5.5cm 5cm 6cm},clip]{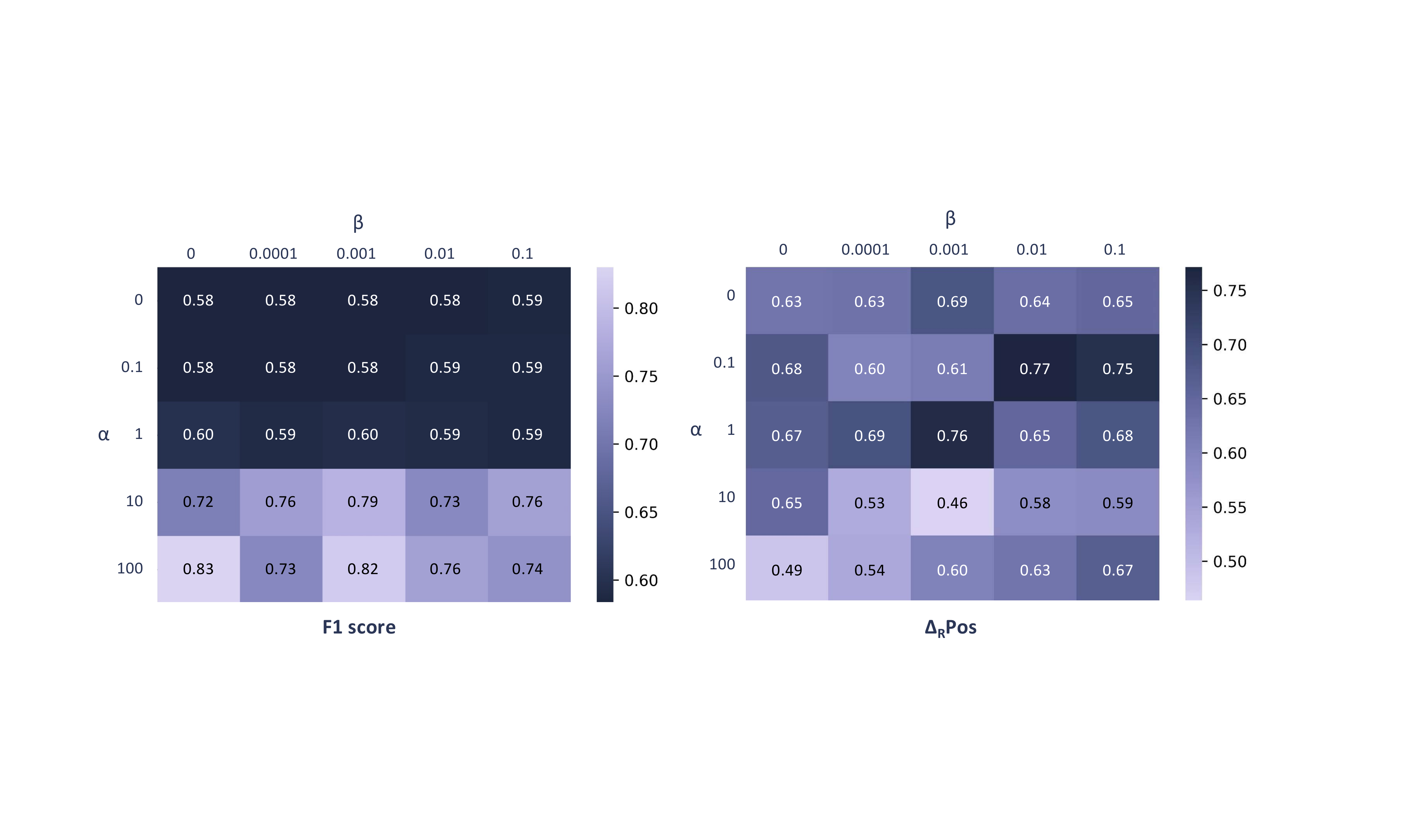}
    \caption{F1 score for per scan CAC detection and per scan absolute relative difference for concordant positive scans on the validation set}
    \label{fig:abl_tabels}
\end{figure}

The adjusted CycleGAN for generating the CAC map was trained with the same data set and preprocessing as the slice classification CNN.
The adjusted CycleGAN was trained for 375,000 iterations, after which the learning curves converged, using batches of 4 axial CT slices of 224x224 voxels, which was the maximum that fit in the available GPU memory. Batches were balanced for both the presence of CAC and relative slice position in the heart, using the heart segmentations. The training parameters were determined experimentally through qualitative (visual) and quantitative assessment. Qualitative assessment evaluated the appearance of the generated synthetic $\hat{noCAC}$ and synthetic $\hat{CAC}$ images, and it led to setting the learning rate to 0.0001 and $\lambda$ to 10\cite{zhu2017unpaired}. Thereafter, quantitative assessment was used to define parameters $\alpha$ and $\beta$ through evaluation of the CAC detection (F1-score) and CAC reproducibility (absolute relative difference) on $RTP_{val}$. Figure \ref{fig:abl_tabels} shows the evaluated parameters and the obtained results. Based on these results, $\alpha = 10$ and $\beta = 0.001$ were used in further experiments. Finally, the predicted CAC maps were resampled to the original in-plane resolution prior to evaluation.

All experiments were performed using Python with Pytorch\cite{pytorch} on an Intel Xeon E5-1620 3.60 GHz CPU with an NVIDIA Titan X GPU. The complete analysis takes on average about 2 minutes per scan.

\begin{figure}[h]
\centering
\includegraphics[height=9cm,trim={0.2cm 10cm 23cm 3cm},clip]{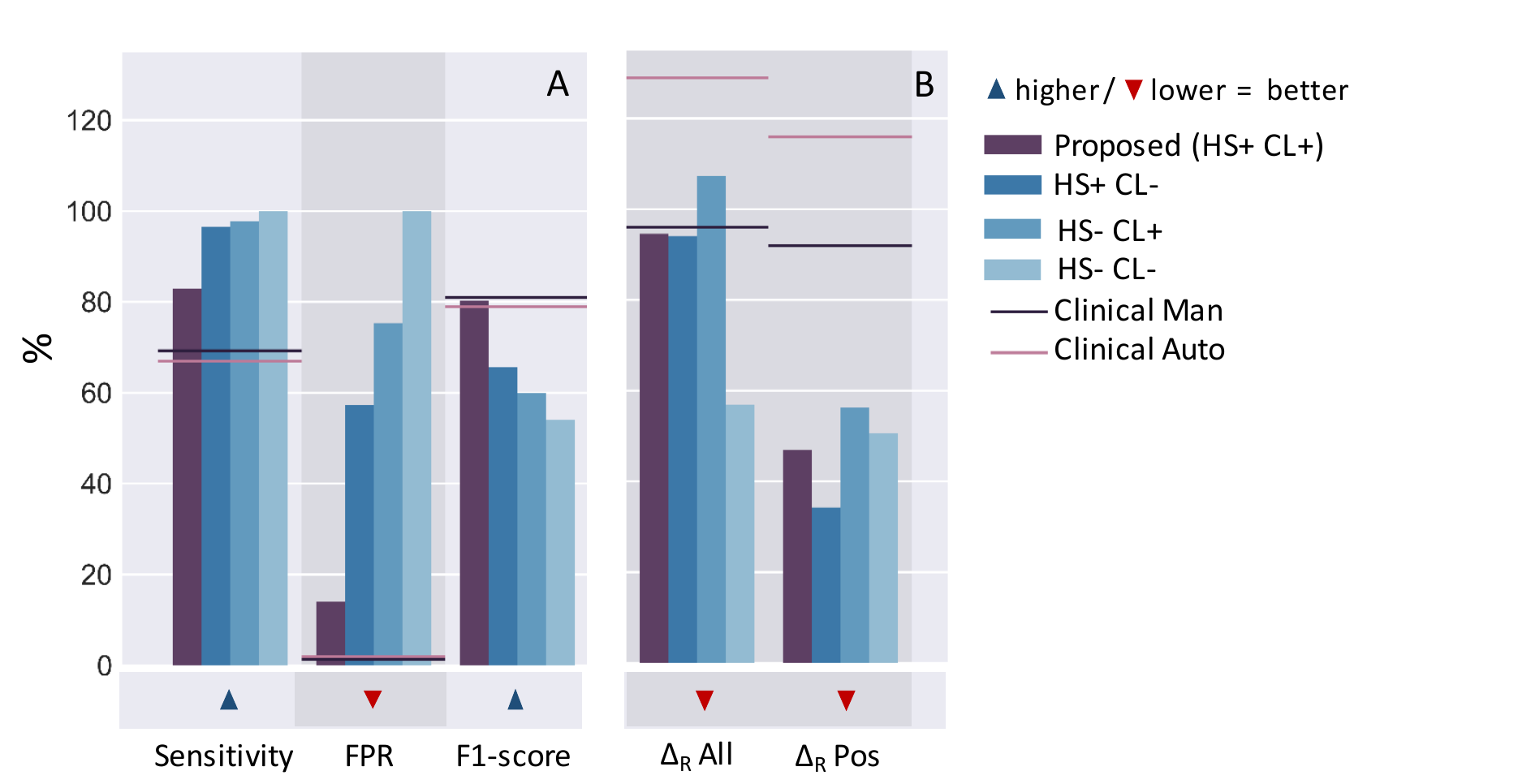}
\caption{The results of the ablation study for the method with and without heart segmentation (HS -/+) and with and without slice classification (CL -/+). We evaluated the detection performance (A) with sensitivity, false positive rate (FPR) and F1-score. We evaluated the interscan reproducibility (B) with absolute relative difference in CAC mass for all pairs ($\Delta_R$ All) and concordant positive pairs ($\Delta_R$ Pos). For comparison, the performance of clinical manual and clinical automatic calcium scoring are shown with horizontal lines.}
\label{fig:ablation}
\end{figure}

\begin{figure*}[h]
\centering
\includegraphics[height=10.8cm,trim={0.8cm 0.5cm 4cm 0.5cm},clip]{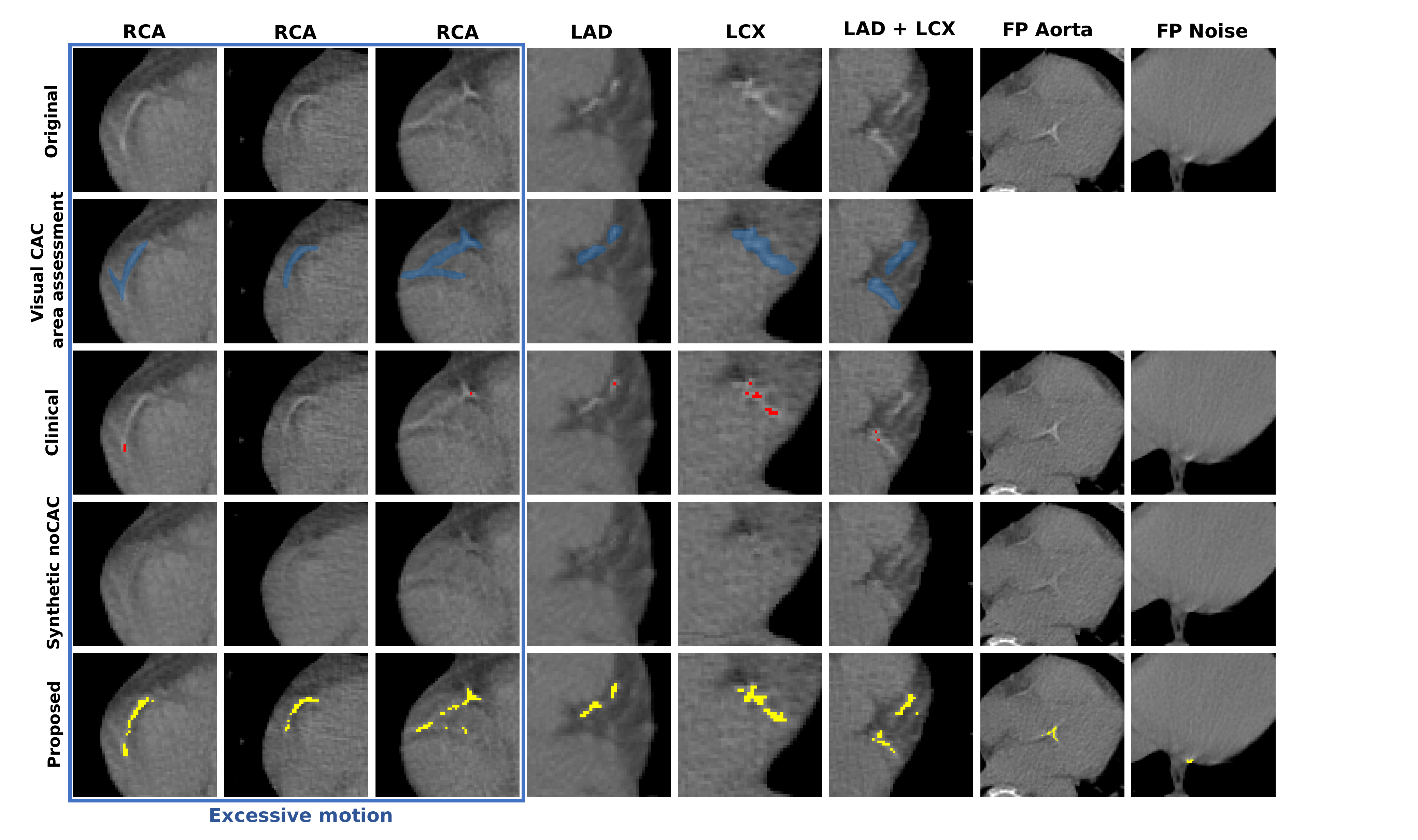}
\caption{Examples of lesions, visual assessment of the CAC lesion area, calcium segmentation using the clinical definition, corresponding synthetic images without CAC and segmentation of CAC with the proposed method and following the clinical protocol. Lesions with excessive motion are indicated with the blue box. }
\label{fig:examples}
\end{figure*}

\subsection{CAC detection per scan}
In the clinic CAC is routinely detected on a scan-level and used for CVD risk prediction. To provide insight in the detection performance of visible CAC on a scan-level, Figure~\ref{fig:examples} illustrates the results and Figure \ref{fig:ablation}A (proposed) shows the sensitivity, rate of false positive scans (FPR) and F1-score in $RTP_{test1}$ (Table \ref{tab:data}). The proposed method achieved a sensitivity for visible CAC of 0.83, with an FPR of 0.14, meaning 21 out of 150 scans without visible CAC, and an F1-score of 0.80. In comparison, the sensitivity was 0.69 and 0.67 for manual and automatic clinical calcium scoring, respectively. The F1-score of the clinical calcium scoring methods was comparable with our method, due to a low FPR.
Among our false positive cases, in 10 cases a lesion in the aorta was erroneously detected (Figure \ref{fig:examples}, FP Aorta), among which 6 were located in the ostia of the coronary arteries. In 9 scans noise was detected as CAC by the proposed method. In 5 of these 9 scans, the false positive lesion was located near the edge of the heart segmentation (Figure \ref{fig:examples}, FP Noise).

As is shown in Figure \ref{fig:examples}, the proposed method also segments the parts of the calcium lesions that remain below the threshold. Particularly in lesions severely affected by motion, e.g. commonly lesions in the RCA, the difference between the proposed method and manual clinical calcium scoring is especially pronounced.

\begin{table}[h]
\footnotesize
\centering
\caption{Per artery detection performance (A) and interscan reproducibility (B) of clinical manual (Clinical\textsubscript{M}) and automatic (Clinical\textsubscript{A}) calcium scoring, and the proposed method in $RTP_{test1}$. CAC detection is shown for pairs and for single scans. Reproducibility is evaluated with the absolute relative difference in CAC pseudo-mass in scan-rescan artery pairs for concordant positive artery pairs ($\Delta_R $Pos) as well as all artery pairs ($\Delta_R $All).}
\label{tab:lesionwise}
\begin{threeparttable}
\begin{tabular}{lllllll}
                    &                        & \multicolumn{2}{c}{(A)}                     &  & \multicolumn{2}{c}{(B)}    \\   
                    &                        & \multicolumn{2}{c}{\textbf{Detection}}    &  & \multicolumn{2}{c}{\textbf{Interscan Reproducibility}} \\
                    &                        & \textbf{Single (\%)} & \textbf{Pairs (\%)} &  & \textbf{$\Delta_R $ All}         & \textbf{$\Delta_R $ Pos$\dagger$}        \\ \cline{1-4} \cline{6-7} 
\multicolumn{3}{l}{\textbf{Proposed}}                            &                     &  &                            &                           \\
                    & CAC                    & \textbf{81 (79)}    & \textbf{34 (67)}    &  & \textbf{0.95}              & \textbf{0.55}             \\
                    & LAD                    & \textbf{57 (92)}    & \textbf{26 (84)}    &  & \textbf{0.84}              & \textbf{0.61}             \\
                    & RCA                    & \textbf{16 (57)}     & \textbf{5 (36)}    &  & \textbf{1.25}              & \textbf{0.35}             \\
                    & LCX                    & \textbf{8 (67)}     & \textbf{3 (50)}     &  & 0.99              & \textbf{0.33}                      \\ \cline{1-4} \cline{6-7}
\multicolumn{3}{l}{\textbf{Clinical\textsubscript{M}}} &                                          &  &                            &                           \\
                    & CAC                    & 70 (69)             & 30 (59)             &  & 1.05                       &  0.77                     \\
                    & LAD                    & 53 (85)             & 23 (74)             &  & 1.00                       & 0.75                      \\
                    & RCA                    & 10 (36)              & 4 (29)             &  & 1.42                       & 1.14                      \\
                    & LCX                    & 7 (58)              & 3 (50)              &  & \textbf{0.84}                       & 0.46             \\ \cline{1-4} \cline{6-7} 
\multicolumn{3}{l}{\textbf{Clinical\textsubscript{A}}}   &                                         &  &                            &                           \\
                    & CAC                    & 64 (63)             & 26 (51)             &  & 1.15                       & 0.76                      \\
                    & LAD                    & 48 (77)             & 21 (68)             &  & 1.07                        & 0.81                     \\
                    & RCA                    & 9 (32)             &  2 (14)              &  &  1.63                       &0.74                      \\
                    & LCX                    & 7 (58)              & 3 (50)              &  &  0.86                       & 0.48                     \\ \cline{1-4} \cline{6-7}

\end{tabular}
\begin{tablenotes}
\item[$\dagger$] Pairs in which CAC was detected in the artery in both scans by the respective method.
\end{tablenotes}
\end{threeparttable}

\end{table}

\subsection{CAC detection per artery}
Because the impact of cardiac motion is different per location, we additionally evaluate the detection performance per artery. A subset of 34 concordant scan pairs from $RTP_{test1}$ that contain CAC with per artery CAC annotations ($RTP_{test lesions}$) were used for evaluation. Overall, the proposed method detected CAC in 81 out of 102 arteries with lesions (Table \ref{tab:lesionwise}A, left). Out of 51 lesion pairs, the method detected both lesions in 34 pairs (Table \ref{tab:lesionwise}A, right). In comparison, manual clinical calcium scoring detected 70 out of 102 single lesions, and automatic clinical calcium scoring detected 64 out of 102. Manual clinical calcium scoring detected CAC in both lesions in 30 out of 51 lesion pairs and automatic clinical calcium scoring detected 26 pairs out of 51. The achieved improvement of our method over the clinical methods was largest in the RCA, in which 64\% of the visible lesions remained below the 130 HU threshold due to blurring caused by cardiac motion. This is also illustrated in Figure \ref{fig:examples}, where all lesions are affected by partial volume effect due to low resolution and especially lesions in the RCA are subject to excessive motion artefacts.

\begin{table*}[]
\centering
\footnotesize
\caption{Reproducibility of CAC pseudo mass and Agatston scores for automatic clinical calcium scoring and the proposed method in $RTP_{test2}$}.
\label{tab:big_tes_set}
\begin{tabular}{llllllll}

 & \multicolumn{3}{l}{\textbf{All pairs}}& \multicolumn{4}{l}{\textbf{Concordant positive pairs}}   \\
\textbf{}  & \multicolumn{2}{l}{CAC mass}& AG score & & \multicolumn{2}{l}{CAC mass} & AG score \\
\textbf{1543 pairs}  &  $\Delta_R$ & ICC & ICC & n (\%) &  $\Delta_R$ & ICC & ICC  \\ \hline
\textbf{Proposed}  & 0.88  & 0.97 (0.96-0.97) & 0.96 (0.96-0.96) & 515 (34\%) & 0.44 & 0.96 (0.96-0.97) & 0.95(0.95-0.96)\\
\textbf{Clinical}   & 0.92  & 0.93 (0.92-0.94) & 0.91 (0.91-0.92) & 375 (24\%) & 0.58 & 0.92 (0.90-0.93) & 0.90 (0.88-0.92)\\ \hline
\end{tabular}
\end{table*}

\subsection{Interscan reproducibility of CAC quantification}
To make CAC quantification in non-ECG-synchronized CT scans useful for clinical application the interscan reproducibility ideally should be high. Therefore, the interscan reproducibility of CAC pseudo mass was evaluated in $RTP_{test1}$. 
First, we evaluated the absolute relative difference in CAC pseudo mass in 51 artery pairs using per artery annotations of lesions described in Section \ref{section:data}. The interscan difference was 95\% on average in all artery pairs (Table \ref{tab:lesionwise}B, left). In concordant positive artery pairs, i.e. pairs in which the method detected CAC in both arteries, the interscan difference was 55\% using the proposed method (Table \ref{tab:lesionwise}B, right). Second, we evaluated the reproducibility of total CAC pseudo mass in scan pairs. The interscan difference was 94\% on average in all scan pairs (Figure \ref{fig:ablation}B, left). For concordant positive scan pairs the absolute relative difference was 47\% (Figure \ref{fig:ablation}B, right). The higher absolute relative difference in CAC pseudo mass (i.e. lower reproducibility) in all scan pairs compared to positive pairs, is partly explained by the presence of 21 discordant pairs. In these pairs the method shows a high absolute relative difference, since a CAC lesion is only visible in one of the two scans. Additionally, we compared our results to clinical calcium scoring. Both clinical methods show a lower interscan reproducibility than our proposed method in artery pairs as well as in scan pairs. For manual clinical calcium scoring the interscan difference was 77\% in concordant positive artery pairs and 89\% in concordant positive scan pairs. For automatic clinical calcium scoring the interscan difference was 76\% and 114\% in artery pairs and scan pairs, respectively. For similar reasons as for our method, the relative interscan difference is high when all scan pairs are considered.



\begin{figure}
    \centering
    \includegraphics[height=10 cm,trim={0.3cm 3.5cm 13cm 0.5cm},clip]{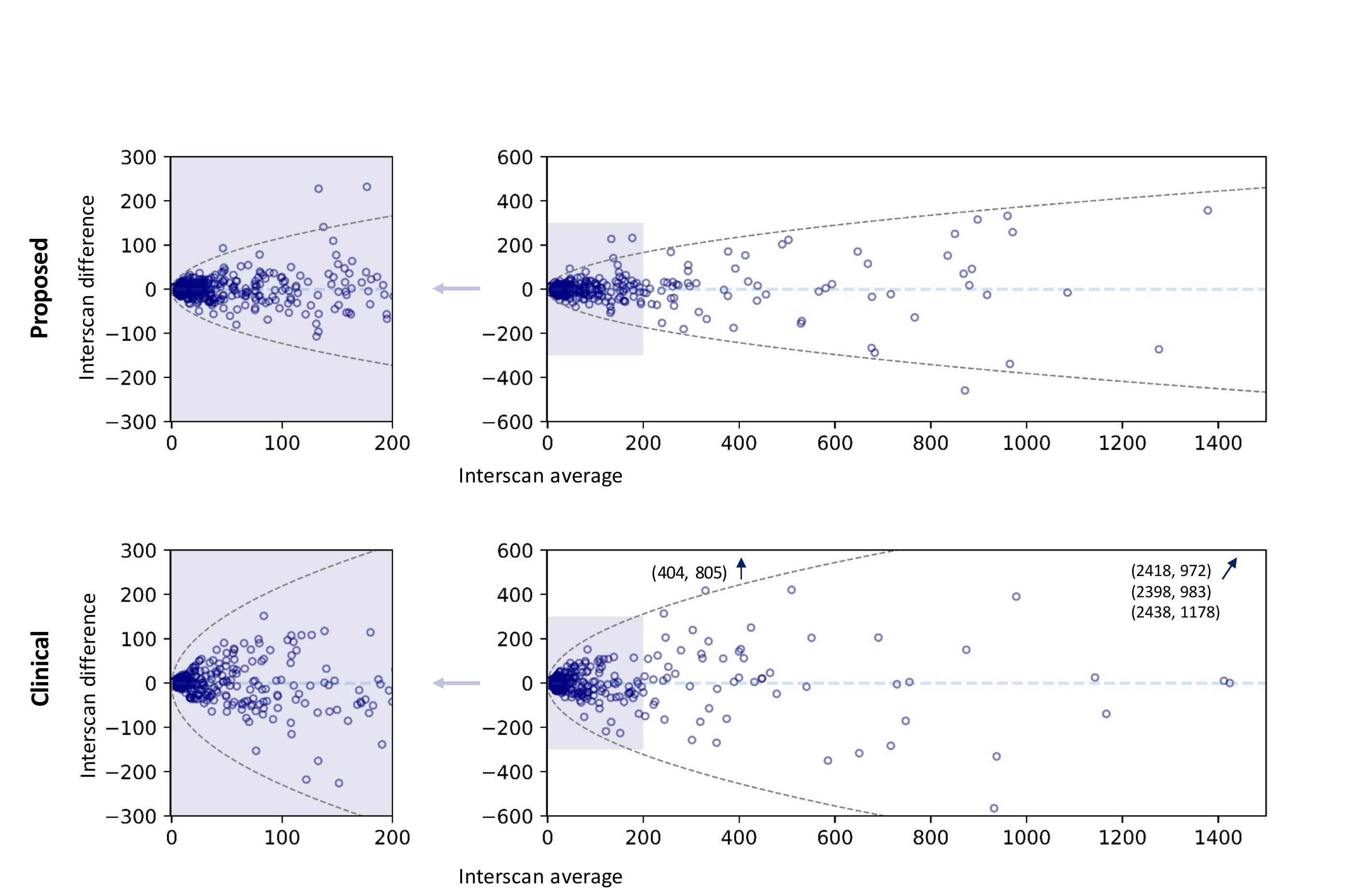}
    \caption{Bland-Altman plots for adjusted Agatston scores of the proposed and automatic clinical method. For comparison purposes the adjusted Agatston scores of the proposed method are scaled to the range of scores of the clinical method. Please note that, since the same linear scaling is used for all scores, this does not influence the agreement. 95\% limits of agreement are indicated by dashed lines. Because the errors tend to increase with increasing CAC, regression for nonuniform differences was used to model the variation of the absolute differences in scan pairs. To calculate the 95\% limits of agreement, the predicted absolute differences were multiplied by $1.96\times(\pi/2)^{1/2}$, because the absolute differences have a half-normal distribution \cite{sevrukov2005serial}.}
    \label{fig:BA_plots}
\end{figure}

Moreover, the reproducibility of the proposed method is evaluated in $RTP_{test2}$, in which manual annotations are not available. Hence, the results of our method are compared with automatic clinical calcium scoring. The results are listed in Table~\ref{tab:big_tes_set}. The relative difference in CAC mass for positive pairs with detected CAC in both scans was 44\%. For all scan pairs this was 88\%. Moreover, the interscan agreement of CAC mass was high with an ICC of 0.96 (95\% CI 0.96-0.97) in positive pairs and 0.97 (95\% CI 0.96-0.97) in all pairs. In comparison, the relative difference was 58\% in concordant positive pairs and 92\% in all pairs when using the automatic clinical method. Moreover, the ICC for the quantified CAC mass was significantly lower for the clinical automatic method than for the proposed method (Table~\ref{tab:big_tes_set}).


\begin{table}[]
\small
\caption{Interscan agreement of risk categorization based on (adjusted) Agatston score in $RTP_{test1}$ and $RTP_{test2}$. The risk categories are I: 0-10, II: 11-100, III: 101-400, IV: $\geq$400.}

\label{tab:risk_categories}
\centering
\begin{threeparttable}[b]
\begin{tabular}{lllll}
 &         & \textbf{Proposed$\dagger$} & \textbf{Clinical\textsubscript{M}} & \textbf{Clinical\textsubscript{A}}  \\ \hline
\multicolumn{5}{l}{\textbf{$RTP_{test1}$}}                                          \\
 & $\kappa$   & 0.82 (0.71-0.94)  & 0.63 (0.42-0.84)      & 0.64 (0.41-0.87)        \\
 & Accuracy  & 0.92 & 0.92                  & 0.95                               \\ \hline
\multicolumn{5}{l}{\textbf{$RTP_{test2}$}}                                          \\
\multicolumn{5}{l}{All pairs}                                                    \\
 & $\kappa$  & 0.84 (0.82-0.87)  & -                     & 0.81 (0.78-0.84)         \\
 & Accuracy   & 0.90 & -                     & 0.92                               \\
\multicolumn{5}{l}{Concordant positive pairs$\ddagger$}                                    \\
 & $\kappa$   & 0.77 (0.72-0.82) & -                     & 0.67 (0.61-0.72)         \\
 & Accuracy & 0.79  & -                     & 0.68                               \\ \hline
\end{tabular}
\begin{tablenotes}
\item[$\dagger$] For comparison purposes, the adjusted Agatston scores obtained with the proposed method were linearly scaled to the range of the scores obtained with automatic clinical (Clinical\textsubscript{A}) calcium scoring, using the mean score of both methods.
\item[$\ddagger$] Pairs in which the Clinical\textsubscript{A} method found CAC in both scans.
\end{tablenotes}
\end{threeparttable}
\end{table}

The proposed method achieved an interscan agreement for adjusted Agatston scores with an ICC of 0.96 (95\% CI 0.96-0.96). This is significantly better than for the automatic clinical method that achieved an ICC of 0.91 (95\% CI 0.91-0.92). This improvement is clearly visible in the Bland-Altman plots (Figure \ref{fig:BA_plots}), where the 95\% limits of agreement are narrower for the proposed method than for automatic clinical calcium scoring. 


\subsection{Ablation Study}
To investigate the benefits of prior heart segmentation and slice classification, we evaluated performance using the following settings: 1) using the settings as proposed, 2) the proposed method without slice classification, where all CT slices of the heart were analysed, 3) the proposed method without heart segmentation, 4) the proposed method without heart segmentation and without slice classification. When heart segmentation was not used, we standardized the field of view over the data sets to include the heart and chest wall. CAC detection, FPR and absolute relative interscan difference of the quantified CAC mass was evaluated on $RTP_{test1}$.
Figure \ref{fig:ablation} shows the results. The proposed method achieved the best performance, with an F1-score of 0.80 and an absolute relative difference in CAC mass of 47\%. While both approaches without slice classification show a high sensitivity, the false positive rate is high: with settings 2 the method found false positive lesions in all negative scans and with settings 4 in 57\% of the negative scans. The effect of defining the region of interest by heart segmentation is not only reflected in a better detection, but also in a higher interscan reproducibility of CAC mass, i.e. lower absolute relative difference. 


\subsection{Risk category assignment}
Because CVD risk stratification based on the Agatston score is clinically relevant, we evaluate the interscan reproducibility of risk categorization. The interscan agreement of CVD risk categories is shown in Table \ref{tab:risk_categories}. The agreement in $RTP_{test1}$ was higher using scores derived with the proposed ($\kappa$ = 0.82) method than with manual of automatic clinical calcium scoring, with $\kappa$ of 0.63 and 0.64 respectively. In concordant positive pairs of $RTP_{test2}$ the $\kappa$ was 0.77 for the proposed method. In contrast, for automatic clinical calcium scoring this was 0.67.

\section{Discussion}
We presented a method for calcium scoring that does not depend on the standard clinically used definition of CAC that applies an intensity value threshold of 130 HU for segmentation of CAC lesions. Instead, our method defines CAC as the difference between an image visibly containing CAC and a healthy tissue image. Hence, the method can identify CAC - complete or parts of lesions - that may remain undetected by the clinical threshold due to e.g. partial volume effect in small lesions or blurring due to cardiac motion. We achieve this by separating an image containing CAC into an image without CAC and an image only containing CAC. For this we exploit a CycleGAN that translates images between the \textit{CAC} and \textit{noCAC} domain. We have shown that segmentation of lesions using our method enables increased interscan reproducibility of CAC quantification, compared to clinically used manual and automatic CAC scoring. Increased interscan reproducibility may lead to more reliable risk estimation and enable longitudinal studies in non-ECG-synchronized CT.

Previous research shows that interscan reproducibility in non-ECG synchronized CT scans is lower than in dedicated cardiac CT~\cite{detrano2005coronary, mao2001effect, hoffmann2006evidence, van2003coronary}. Because of limitations in the acquisition of radiotherapy treatment planning CTs, our method does not exceed the in literature reported reproducibility in \mbox{dedicated} cardiac CT. However, CAC scoring is increasingly performed in non-dedicated CT scans, hence, the need for more reproducible CAC quantification is growing, especially for quantification in challenging scans like the ones used in this work. Our proposed method shows potential for increasing the interscan reproducibility in non-ECG synchronized scans to the level of dedicated cardiac CT.



In contrast to clinical methods, the proposed method is able to detect visible CAC lesions that remain below the threshold. Owing to a more accurate segmentation that includes pairs of lesions below the threshold, our method outperforms clinical CAC scoring methods in interscan reproducibility of CAC quantification. The results of per artery evaluation show that this improvement was especially pronounced for CAC in the RCA. This is probably due to the ability of the proposed method to detect parts of lesions that are heavily affected by motion artefacts, which occurs more for the lesions in the RCA than in the LAD and LCX.


Errors were made in a number of scans. False positive detections mostly consisted of noise in the proximity of an artery or other types of calcifications than CAC, such as calcifications in the aorta or cardiac valves. Despite noise augmentation during training and post processing, the CAC map contained noise in a few cases. Such errors were often located close to the apex and near the RCA, where the noise level was most severe. Noise reduction strategies described in e.g. Wolterink et al.~\cite{wolterink2017generative} could offer a solution. False positives in the aorta or valves were often close to the ostia of the coronaries or occurred in the mitral valve. Since in scans without ECG triggering it is often difficult to distinguish CAC from calcifications in the thoracic aorta or mitral valve calcifications, these type of errors are also not uncommon for human observers~\cite{van2014cardiac} and are also present in other automatic methods \cite{vanvelzen2020deep, niko, devos2019direct}. False negative lesions were typically small with very low voxel intensities, making them difficult to distinguish from noise and soft tissue, also for experts. 


Similarly to our preliminary work~\cite{vanvelzen2020coronary}, the method without heart segmentation and slice classification suffered from a large amount of false positives in the CAC map, which made fully automatic quantification of CAC infeasible. In our ablation experiment, we showed that adding slice classification with respect to CAC presence and heart segmentation solves this issue. Using classification to identify the slices containing CAC, prevents false positive errors in slices that do not contain CAC. By adding heart segmentation, we prevent anatomical structures outside the heart in the data from influencing the CAC segmentation, resulting in more reproducible quantification. Likewise, in previous methods for CAC quantification in scans with high noise levels, a false positive reduction step was used to decrease the number of false positive findings\cite{vanvelzen2020deep, niko}. Possibly, if a very large amount of training data would be available, the method would be able to learn to avoid these errors. However, in the medical field this is often challenging to obtain. In this work, heart segmentation and slice classification are performed independently. Although they are relatively straightforward, future work will investigate merging these steps into a single network. 

The heart segmentation network showed good performance in terms of Dice and asymmetric surface distance. However, a relatively high Hausdorff distance was found. Visual evaluation revealed that errors causing high Hausdorff distance were mostly located in the basal slices of the heart. This is probably due to high interobserver variability in reference segmentations regarding the difficult to outline superior boundaries of the heart. Given that the segmented superior boundaries are typically well above the ostia of the coronaries, we expect the influence of variability in the superior boundaries of the heart on CAC classification and quantification to be minimal.

During experiments the impact of the different loss terms for training the Calcium CycleGAN on the CAC detection and reproducibility was evaluated using relatively large differences in the values of the tested parameters. Possibly a performance gain can be achieved by further optimizing $\alpha$ and $\beta$. Nevertheless, given the obtained results (Figure \ref{fig:abl_tabels}), we expect the benefit would be limited.

Other approaches for increasing the reproducibility of CAC quantification have previously been proposed\cite{groen2009threshold, dehmeshki2007volumetric, vsprem2018coronary,song2019improved,saur2009accuratum}. Since the true amount of CAC is not available for patients, often a phantom is used for development and evaluation of the method. This makes direct comparison of our method infeasible. \v{S}prem et al.\cite{vsprem2018coronary} additionally evaluated their approach in a set of 293 subject CT scans and showed an improved reproducibility of CAC volume between different CT images. However, different reconstructions of the same acquisition were used, excluding interscan differences, in contrast to two separate acquisitions used in our study. Moreover, in contrast to our approach a major drawback of previously proposed methods is the use of the standard 130 HU intensity threshold for detection of CAC prior to partial volume correction, inevitably missing lesions below the threshold.


Risk categorization of subjects is typically done using total CAC Agatston scores. However, the proposed method also includes parts of lesions below the clinical threshold, and thus, the Agatston score tends to be (substantially) higher than for the clinical methods. Therefore, assigning risk categories based on the Agatston score derived with the proposed method would necessarily lead to different definitions of CVD risk categories. In order to estimate the performance of risk categorization of the proposed method, the obtained Agatston scores were scaled to the range of clinical Agatston scores. The results showed an improved agreement for the proposed method, compared to automatic clinical calcium scoring. However, to establish the clinical relevance, further research relating the adjusted Agatston scores with CHD events is warranted. 

Future research on evaluation of the proposed method in non-ECG-synchronized against calcium scoring in dedicated calcium scoring CT is needed. However, clinical calcium scoring in dedicated cardiac CT also suffers from low reproducibility\cite{detrano2005coronary, mao2001effect, hoffmann2006evidence, van2003coronary}. A non-invasive imaging-based reference that indicates the true amount of CAC does not exist. Hence, to define the clinical relevance, future research investigating whether more reproducible CAC quantification with the proposed method leads to better prediction of CVD events is warranted. However, since CVD related outcomes were not available, investigating the relation to CVD events is out of scope of this work.

\section{Conclusion}
We have proposed a method for detection and quantification of CAC in non ECG-synchronized chest CT scans that that does not require the application of the clinically used intensity level threshold. Hence, the method allows detection of visible CAC that may not exceed the standardly used 130 HU threshold due to e.g. partial volume effect or blurring due to cardiac motion. This leads to more reproducible quantification of CAC and increased interscan reproducibility of risk categorization compared to the current clinical method.


%



\section*{Disclosures}
The authors declare no competing interests.

\section*{Acknowledgment}
The authors gratefully acknowledge the Dutch Cancer Society for the financial support (NCT03206333). The authors thank the National Cancer Institute for access to NCI's data collected by the National Lung Screening Trial. The statements contained herein are solely those of the authors and do not represent or imply concurrence or endorsement by NCI.

\bibliography{report}   
\bibliographystyle{spiebib}   





\end{spacing}
\end{document}